\documentclass[10pt,a4paper,oneside]{article}

\usepackage{amssymb}
\usepackage{amsmath}
\usepackage{amsthm}

\newtheorem{theorem}{Theorem}[section]
\newtheorem{lemma}{Lemma}[section]
\newtheorem{conjecture}{Conjecture}[section]

\newcommand{\MN}{\mathcal{M}^{+}_{n}}
\newcommand{\TTMN}{\mathcal{M}_{n}}
\newcommand{\MQ}{\mathcal{Q}^{+}_{n}}

\newcommand{\Tr}{\mathop{\textrm{Tr}}\nolimits}
\newcommand{\Scal}{\mathop{\textrm{Scal}}\nolimits}
\newcommand{\LIN}{\mathop{\textrm{LIN}}\nolimits}
\newcommand{\la}{\lambda}
\newcommand{\va}{\varphi}

\begin{document}

\title{
  On the monotonicity conjecture for the curvature of the Kubo-Mori metric
  \thanks{keywords: state space, Kubo-Mori metric,
                   monotone statistical metric, scalar curvature;
          MSC: 53C20, 81Q99}}
\author{Attila Andai\thanks{andaia@math.bme.hu}\\
  Department for Mathematical Analysis, \\
  Budapest University of Technology and Economics,\\
  H-1521 Budapest XI. Sztoczek u. 2, Hungary}
\date{October 30, 2003}

\maketitle

\begin{abstract}
The canonical correlation or Kubo-Mori scalar product on the state space
  of a finite quantum system is a natural generalization of the classical
  Fisher metric.
This metric is induced by the von Neumann entropy or the
  relative entropy of the quantum mechanical states.
An important conjecture of Petz that the scalar curvature of the state
  space with Kubo-Mori scalar product as Riemannian metric is monotone with
  respect to the majorisation relation of states: the scalar curvature is
  increases if one goes to more mixed states.
We give an appropriate grouping for the summands in the expression
for the scalar
  curvature.
The conjecture will follows from the monotonicity of the summands.
We prove the monotonicity for some of these summands
  and we give numerical evidences that the remaining terms are monotone too.
Note that the real density matrices form a submanifold of the complex density
matrices. We prove that if Petz's conjecture true for complex density
matrices then it is true for real density matrices too.
\end{abstract}

\section{Introduction}

The state space of a finite quantum system can be endowed with a
differentiable
  structure \cite{Pet1}.
The canonical correlation defines a Riemannian structure on it.
There is strong connection between the scalar curvature of this
  manifold at a given state and statistical distinguishability
  and uncertainty of the state \cite{Pet2}.
Roughly speaking the scalar curvature measures the average statistical
  uncertainty.
This idea comes from a series expansion of the volume of the geodesic ball.
If $D_{0}$ is a given point in the state space and $B_{r}(D_{0})$ is
  the geodesic ball with center $D_{0}$ and radius $r$ then the volume of
  this ball is given by
\begin{equation}
V(B_{r}(D_{0}))=\frac{\sqrt{\pi^{n}}r^{n}}{\Gamma\left(\frac{n}{2}+1\right)}
  \cdot\left(1-\frac{\Scal(D_{0})}{6(n+2)}\cdot r^{2}+ O(r^{4}) \right)
\end{equation}
  where $\Scal(D_{0})$ is the scalar curvature at the point $D_{0}$ and
  $n$ is the dimension of the manifold.
There was given an explicit formula for the scalar curvature of these
  manifolds, for example in \cite{Ditt1,Ditt2,MicPet,PetSud}.
Physically it is reasonable to expect that the most mixed
  states are less distinguishable from the neighboring ones
  than the less mixed states, for details see \cite{Pet1,Raj}.
It means mathematically that the scalar curvature
  of physically relevant Riemann structures should have monotonicity
  property in the sense that if $D_{1}$ is more mixed than $D_{2}$ then
  $\Scal(D_{2})$ should be less then $\Scal(D_{1})$.
This was conjectured first by Petz \cite{Pet1} and it was verified for
  $2\times 2$ matrices in \cite{Pet1}.
There were some numerical simulations showing that the conjecture
  is true but no mathematical proof for that.
The aim of the paper is to give an appropriate grouping for the
summands in
  the expression for the scalar curvature and prove that some of these summands
  are monotone with respect to the majorisation.

The paper is organized as follows.
In Section 2 definition, properties and equivalent forms of majorisation
  are given and the conjecture is formulated.
In Section 3 the Kubo-Mori Riemannian inner product is defined and the curvature
  formula is given for real and complex density matrix spaces.
In Section 4 we rewrite the scalar curvature formula as the sum of five
  terms. Four of them correspond with the conjecture as
  elementary, but brutal computations show. Further remarks and a lemma about
  the conjecture is presented as well.
In Section 5 we prove that if the scalar curvature is monotone with respect to the
  majorisation on the
  space of complex density matrices then it is monotone on the space of
  real ones.

\section{Majorisation}

The state space of a finite quantum system is the set of positive
  semidefinite $n\times n$ matrices of trace 1.
Such matrices are often called density matrices.
Let $\MQ$ denote the space of invertible density matrices.
The state $A$ is called majorised by the state $B$,
  denoted by $A\prec B$ if the following hold
  for their decreasingly ordered set of eigenvalues $(a_{1},\dots,a_{n})$ and
  $(b_{1},\dots,b_{n})$
\begin{equation}
  \sum_{l=1}^{k} a_{l}\leq\sum_{l=1}^{k}b_{l}
\end{equation}
for all $1\leq k\leq n$.
For any $(a_{1},\dots,a_{n})$ eigenvalues
of a density matrix
\[\left(\frac{1}{n},\dots,\frac{1}{n}\right)\prec
  (a_{1},\dots,a_{n})\prec (1,0,\dots,0).\]
The majorisation occurs naturally in various contexts. For example,
  in statistical quantum mechanics the relation $A\prec B$ is interpreted
  to mean that the state $A$ describes a more mixed or a "more chaotic"
  state than $B$ \cite{AlbUhl}.
A linear map $T$ on $\mathbb{R}^{n}$ a $T$-transform if there exists
  $0\leq t\leq 1$ and indices $k,l$ such that
\begin{equation*}T(x_{1},\dots,x_{n})=(x_{1},\dots,x_{k-1},tx_{k}+(1-t)x_{l},x_{k+1},\dots,
  x_{l-1},(1-t)x_{k}+tx_{l},x_{l+1},\dots,x_{n}).
\end{equation*}
Note that if the density matrix $A$ has eigenvalues
  $(a_{1},\dots,a_{n})$ and the
  density matrix $B$ has eigenvalues $T(a_{1},\dots,a_{n})$ then
  $B\prec A$.

From every self-adjoint $H$ operator and positive parameter $\beta$
  one can set a state
\begin{equation}
 R_{H}(\beta)=\frac{e^{-\beta H}}{\Tr e^{-\beta H}}
\end{equation}
  which is called the Gibbs state at the inverse temperature $\beta$ for
  the Hamiltonian H.
One can check that
  if the difference $H_{1}-H_{2}$ is a multiple of the identity then
  $R_{H_{1}}(\beta)=R_{H_{2}}(\beta)$.
It is known that if $\beta_{1}<\beta_{2}$ then
  $R_{H}(\beta_{1})\prec R_{H}(\beta_{2})$ \cite{OhyPet}.
This last result means that the states are more mixed at higher temperature.

\begin{theorem}\label{th:2.1.}
  Assume that we have two invertible states $A$ and $B$ with decreasingly
  ordered set of eigenvalues $(a_{1},\dots,a_{n})$ and $(b_{1},\dots,b_{n})$.
  The following are equivalent:
\begin{enumerate}
\item The state $A$ is more mixed than $B$. \item One can find a
sequence $(C_{z})_{z=1,\dots,d}$ between them such that
      for all $z=1,\dots,d$: $C_{z}\in\MQ$,
      \[A=C_{1}\prec C_{2}\prec\dots\prec C_{d}=B\]
      holds and the set of eigenvalues of $C_{z}$ and $C_{z-1}$
      is the same except two elements.
\item The set $(a_{1},\dots,a_{n})$ is obtained from $(b_{1},\dots,b_{n})$
      by a finite number of T-transforms.
\item There is a sequence $(G_{z})_{z=1,\dots,d}$ between them such that
      for all $z=1,\dots,d$: $G_{z}\in\MQ$,
      \[A=G_{1}\prec G_{2}\prec\dots\prec G_{d}=B\]
      holds and for all $i=1,\dots,d-1$ there exists a selfadjoint
      operator  $H_{i}$ and positive parameters $\beta_{1,i},\beta_{2,i}$ such that
      $G_{i}=R_{H_{i}}(\beta_{1,i})$ and $G_{i+1}=R_{H_{i}}(\beta_{2,i})$.
\end{enumerate}
\end{theorem}
\begin{proof}
That (1) and (3) are equivalent can be found for example in \cite{Raj}.
The statement (2) is just the reformulation of (3).
The implication (3) $\to$ (2) is trivial and easy to check that
(2) $\to$ (3) implication holds too.
\end{proof}

After this introduction we can formulate correctly the Petz`s conjecture \cite{Pet1}.

\begin{conjecture}
If $D_{1}\prec D_{2}$ then $\Scal(D_{1})>\Scal(D_{2})$
where $\Scal(D)$ denotes the scalar curvature of the state space induced
by the canonical correlation inner product as Riemannian metric at the
point $D$.
\end{conjecture}

\section{Scalar curvature formula}

Let $\MN$ be the space of all complex self-adjoint positive
  definite $n\times n$ matrices and let $\TTMN$ be the real
  vector space of all self-adjoint $n\times n$ matrices.
The space $\MN$ can be endowed with a differentiable structure
\cite{HiaPet}
  and the tangent space $T_{D}\MN$ at $D\in\MN$ can be identified with
  $\TTMN$ \cite{HiaPet}.
One can consider the quantum mechanical state space $\MQ$ as a Riemannian
  submanifold of $\MN$ of codimension 1.

There is a very important functional on the space $\MN$, namely the
  von Neumann entropy
\begin{equation}
  S(D)=-\Tr D\log(D).
\end{equation}
Since this functional is strictly concave \cite{OhyPet,Raj}, the
  second derivative of the entropy
\begin{equation}
  ddS:\MN\to\LIN(\TTMN\times\TTMN,\Bbb R)\quad
    D\mapsto\left( (X,Y)\mapsto -\int_{0}^{\infty}
     \Tr\bigl((D+t)^{-1}X(D+t)^{-1}Y\bigr)  \right)
\end{equation}
  is negative definite.
The Riemannian metric which arises as the negative of the second derivative
  of the entropy
\begin{equation}
  G_{D}(X,Y)=\int_{0}^{\infty}\Tr\bigl((D+t)^{-1}X(D+t)^{-1}Y\bigr)dt
\end{equation}
  is called Kubo-Mori metric.
  This scalar product is an important ingredient of linear response theory
  and often called the canonical correlation of $X$ and $Y$.
\medskip

Let us introduce this metric in a different way.
Recall that Umegaki's relative entropy
\begin{equation}
S(D_{1},D_{2})=\Tr D_{1}(\log D_{1}-\log D_{2})
\end{equation}
of density matrices measures the information between the corresponding states
\cite{OhyPet}.
The partial derivatives of the relative entropy is the Kubo-Mori inner product:
\begin{equation}
\frac{\partial^{2}}{\partial t\partial s}\bigm\vert_{t=s=0}S(D+tX,D+sY)
=G_{D}(X,Y).
\end{equation}
\medskip

Let $D\in\MN$ and choose a basis of $\Bbb R^{n}$ such that
  $D=\sum_{k=1}^{n}\la_{k}E_{kk}$ is diagonal, where $(E_{jk})_{j,k=1,\dots,n}$
  are the usual system of matrix units.
Define the self-adjoint matrices
  \[F_{kl}=E_{kl}+E_{lk}\quad H_{kl}=i\ E_{kl}-i\ E_{lk},\]
  then the $(F_{kl})_{1\leq k<l\leq n}$, $(H_{kl})_{1\leq k<l\leq n}$ and
  $(F_{kk})_{1\leq k\leq n}$ vector system is a basis in $\TTMN$.
Let us define for positive numbers $x,y,z$ the functions
\begin{equation}
  m(x,y)  :=\int_{0}^{\infty}\frac{1}{(x+t)(y+t)}dt \qquad
  m(x,y,z):=\int_{0}^{\infty}\frac{1}{(x+t)(y+t)(z+t)}dt.
\end{equation}
The scalar product of the basis vectors in the tangent space at the point $D$:
\begin{align}
&\text{for}\ 1\leq i<j\leq n,\ 1\leq k<l\leq n:&\
  &\left\lbrace\begin{array}{l}
    G_{D}(H_{ij},H_{kl})=G_{D}(F_{ij},F_{kl})
    =\delta_{ik}\delta_{jl}2m(\la_{i},\la_{j}) \\
    G_{D}(H_{ij},F_{kl})=0
  \end{array} \right. \nonumber\\
&\text{for}\ 1\leq i<j\leq n,\ 1\leq k\leq n:&\
  &G_{D}(H_{ij},F_{kk})=G(F_{ij},F_{kk})=0 \\
&\text{for}\ 1\leq i\leq n,\ 1\leq k\leq n:&\
  &G_{D}(F_{ii},F_{kk})=\delta_{ik}4m(\la_{i},\la_{i})\nonumber
\end{align}
where $\delta_{ik}=1$ if $i=k$ else $\delta_{ik}=0$.
We will use the following properties of the $m$ functions:
assume that $x,y,z$ and $\mu$ are different positive numbers, then
\begin{align}
 &m(x,y)=\frac{\log x-\log y}{x-y},         &\
 &m(x,y,z)=\frac{m(x,z)-m(y,z)}{x-y},       &\
 &m(x,x)=\frac{1}{x}\quad m(x,x,x)=\frac{1}{2x^{2}},\label{eq:mid}   \\
 &m(x,y)=\frac{1}{\mu}m(\mu x,\mu y),       &\
 &m(x,x,y)=\frac{m(x,y)-\frac{1}{x}}{x-y},  &\
 &m(x,y,z)=\frac{1}{\mu^{2}}m(\mu x,\mu y,\mu z).\nonumber
\end{align}
The following identity will be used several times
\begin{equation}
\frac{1}{m(x,y)}\left(\frac{m(x,x,y)}{m(x,x)}+\frac{m(x,y,y)}{m(y,y)} \right)=1.
\end{equation}
Let us define two functions
\begin{eqnarray}\label{eq:ing}
&\va(x,y,z)&=\frac{1}{2}\frac{m(x,y,z)^{2}}{m(x,y)m(y,z)m(z,x)}-
  \frac{m(y,y,x)m(y,y,z)}{m(y,x)m(y,y)m(y,z)}, \\
&v(x,y)&=\frac{1}{2}\frac{m(x,x,y)^{2}}{m(x,y)^{2}m(x,x)}-
  \frac{m(x,x,y)m(y,y,x)}{m(x,y)^{2}m(y,y)}.\nonumber
\end{eqnarray}
From the definition and the identities (\ref{eq:mid}) one can derive
  scaling properties
\begin{equation}
\va(\mu x,\mu y,\mu z)=\frac{1}{\mu}\va(x,y,z)\quad
   v(\mu x,\mu y)=\frac{1}{\mu}v(x,y).
\end{equation}
\medskip

The space $\MQ$ is a one codimensional submanifold of $\MN$.
The tangent space of $\MQ$ is the set of self-adjoint traceless
matrices.
The scalar curvature of this submanifold can be computed using the
Gauss equation \cite{Ama,GalHul}.
\medskip

The scalar curvature the space of complex density matrices at a
given matrix $D$ with eigenvalues $\lambda_{1},\dots,\lambda_{n}$
is
\begin{equation}\label{eq:scal}
  \Scal(D)=\sum^{n}_{\begin{array}{c}
            \scriptstyle j,k,l=1\\
            \scriptstyle \vert\lbrace j,k,l\rbrace\vert>1
            \end{array}}
  \varphi(\la_{j},\la_{k},\la_{l}),
\end{equation}
where $\vert\lbrace j,k,l\rbrace\vert>1$ means that the indices $i,j$ and $k$ are not
  equal \cite{Ditt1}, and that in the space of real density matrices is
\begin{equation}\label{eq:scalr}
\Scal_{\Bbb R}{D}=\frac{1}{4}\sum^{n}_{\begin{array}{c}
            \scriptstyle j,k,l=1\\
            \scriptstyle \vert\lbrace j,k,l\rbrace\vert>1
            \end{array}}
   \varphi(\la_{j},\la_{k},\la_{l})+
  \frac{1}{4}\sum^{n}_{k,l=1}v(\la_{k},\la_{l}).
\end{equation}
\cite{MicPet}. Note that the scalar curvature at a given point depends
only on the eigenvalues of the density matrix.

\section{About monotonicity of the scalar curvature}

Using Theorem {\ref{th:2.1.}} the monotonicity conjecture will
follows from the inequality
\[\Scal(A)=\Scal(C_{1})\leq \Scal(C_{2})\leq\dots\leq \Scal(C_{d})=\Scal(B).\]

Let $A$ and $B$ be two states, such that $A\prec B$ and the decreasingly ordered
  set of eigenvalues of $A$ and $B$ be the same except two  elements.
In this case their eigenvalues can be written in the form
  $(\la_{1},\dots,a,\dots,b,\dots,\la_{n})$ and
  $(\la_{1},\dots,a-x,\dots,b+x,\dots,\la_{n})$, where $x\leq\frac{a-b}{2}$.

If one computes the scalar curvature using equation (\ref{eq:scal}) the summands
  can be grouped:
\begin{enumerate}
\item Summands where only $a$ and $b$ appear
\begin{equation}\label{eq:alpha}
    \alpha(a,b):=2\va(a,a,b)+2\va(b,b,a)+\va(a,b,a)+\va(b,a,b).
\end{equation}
\item Summands where $a,b$ and another eigenvalue appears
\begin{eqnarray}
 \label{eq:beta1}&\beta_{1,k}(a,b):=2\va(a,a,\la_{k})+2\va(b,b,\la_{k})+
                      \va(a,\la_{k},a)+ \va(b,\la_{k},b)&\\
 \label{eq:beta2}&\beta_{2,k}(a,b):=2\va(a,b,\la_{k})+2\va(b,a,\la_{k})+
                      \va(a,\la_{k},b)+ \va(b,\la_{k},a).&
\end{eqnarray}
\item Summands where $a$ or $b$ appear only once
\begin{equation}\label{eq:gamma}
\begin{split}
 \gamma_{kl}(a,b):=
    &\va(a,\la_{k},\la_{l})+\va(b,\la_{k},\la_{l})+\va(\la_{k},a,\la_{l})\\
    &+\va(\la_{k},b,\la_{l})+\va(\la_{k},\la_{l},a)+\va(\la_{k},\la_{l},b).
\end{split}
\end{equation}
\item Summands without $a$ and $b$
  \[\delta_{jkl}:=\va(\la_{j},\la_{k},\la_{l}).\]
\end{enumerate}

After this grouping the curvature formula is
\begin{equation}\label{eq:scalkif}
\Scal(D)=\alpha(a,b)
  +\sum_{k=1}^{n}\bigl(\beta_{1,k}(a,b)+\beta_{2,k}(a,b)\bigr)
  +\sum_{k,l=1}^{n}\gamma_{kl}(a,b)
  +\sum^{n}_{\begin{array}{c}
            \scriptstyle j,k,l=1\\
            \scriptstyle \vert\lbrace j,k,l\rbrace\vert>1
            \end{array}}\delta_{jkl}.
\end{equation}

One possible way to prove the monotonicity conjecture is to show that every summand in
the previous formula is monotone with respect to the majorisation.
\medskip

Dittmann used the symmetrization of function $\va$ and the previous
grouping of the scalar curvature in \cite{Ditt1}.
Numerical tests were confirmed by Dittmann about Petz's monotonicity
conjecture and suggestive 3D-plots were given in \cite{Ditt1}.
\medskip

\begin{theorem}\label{th:4.1.} Part $\alpha$ of the scalar curvature is
monotone, that is for given $a>b$ positive numbers the function
$\alpha(a-x,b+x)$
is strictly increasing in the variable
$x\in\left\lbrack 0,\frac{a-b}{2}\right\rbrack$.
\end{theorem}
\begin{proof}
From the identities (\ref{eq:mid}),(\ref{eq:ing}) it follows, that
\[\varphi(a,b,a)+\varphi(b,a,b)=\frac{-1}{2(a+b)}\frac{1+b/a}{m(1,b/a)^{2}}
\bigl(m(1,1,b/a)^{2}+(b/a)m(b/a,b/a,1)^{2} \bigr)\]
and
\[2\va(a,a,b)+2\va(b,b,a)=\frac{-1}{a+b}(1+b/a)^{2}
  \frac{m(1,1,b/a)m(b/a,b/a,1)}{n(1,b/a)^{2}}.\]
From these formulas and the special values of the functions $m$
  in (\ref{eq:mid}) we obtain
\[\alpha(a-x,b+x)=\frac{-1}{a+b}\cdot\left(
  -\frac{1}{2}\frac{(1+c(x))^{2}}{(1-c(x))^{2}}
  +\frac{(1+c(x))(1+c(x)^{2})}{c(x)(c(x)-1)\log c(x)}
  -\frac{1}{2}\frac{(1+c(x))^{2}}{c(x)\log^{2}c(x)} \right),\]
where
\[c(x)=\frac{b+x}{a-x}.\]
Since the function $c(x)$ increasing, to prove the theorem it
is enough to show that the function
\[\tau(c)=-\frac{1}{2}\frac{(1+c)^{2}}{(1-c)^{2}}
  +\frac{(1+c)(1+c^{2})}{c(c-1)\log c}
  -\frac{1}{2}\frac{(1+c)^{2}}{c\log^{2}c}\]
is decreasing on the interval I=$\lbrack 0,1 \rbrack$. To show
that the function
\[\tau'(c)= \frac{(c+1)^{2}}{c^{2}\log^{3} c}
           -\frac{(c+1)(3c^{2}-2c+3)}{c^{2}(c-1)\log^{2} c}
           +\frac{c^{4}-2c^{3}-2c^{2}-2c+1}{c^{2}(c-1)^{2}\log c}
           +\frac{2c+2}{(c-1)^{3}}\]
is negative or equivalently  the function
\[\tau_{1}(c)=(c-1)^{3}\log^{3} c\cdot\tau'(c)\]
is negative on the interval $I$, enough to prove that
\smallskip

(1)  $\tau_{1}'(c)>0$ on $I$ and $\lim_{c\to 1}\tau_{1}(c)=0$.
\smallskip

The limit can be easily checked and the first part will follow from the
statement:
\smallskip

(2) $\tau_{2}(c)>0$ on $I$ and $\lim_{c\to 1}\tau_{1}'(c)=0$,
\smallskip

where
\[\tau_{2}(c)=\frac{c^{4}}{(c-1)(c^{2}-c+1)(c^{2}+c+1)}\cdot \tau_{1}'(c).\]

Substituting $\tau_{1}'$ into the previous formula:
\[\tau_{2}(c)=6\log^{2}c
             +\frac{(c-1)(c+1)(c^{2}-12c+1)}{(c^{2}+c+1)(c^{2}-c+1)}\cdot\log c
             +\frac{(c+1)^{2}(c-1)^{2}}{2(c^{2}+c+1)^{2}(c^{2}-c+1)^{2}}.\]
The limit in the statement (2) again can be checked and the positivity of
$\tau_{2}(c)$ will follow from the next statement
\smallskip

(3) $\tau_{3}(c)<0$ on $I$ and $\lim_{c\to 1}\tau_{2}(c)=0$,
\smallskip

where
\[\tau_{3}(c)=c(c^{2}+c+1)^{2}(c^{2}-c+1)^{2}\cdot\tau_{2}'(c).\]
Computation shows that
\begin{equation*}
\begin{split}
\tau_{3}(c)=&2(6c^{8}+6c^{7}+13c^{6}-24c^{5}+22c^{4}-24c^{3}+13c^{2}+6c+6)\log c\\
            &+(c^{8}-12c^{7}+4c^{6}-4c^{2}+12c-1).
\end{split}
\end{equation*}
The limit in the statement (3) again can be checked and the negativity of
$\tau_{3}(c)$ will follow from the statement
\smallskip

(4) $\tau_{3}'(c)>0$ on $I$ and $\lim_{c\to 1}\tau_{3}(c)=0$.
\smallskip

Easy to check that the limit condition fulfils in the previous statement,
and the inequality is the following:
\begin{equation*}
\begin{split}
 &4(24c^{7}+21c^{6}+39c^{5}-60c^{4}+44c^{3}-36c^{2}+13c+3)\log c\\
 &+2\frac{(c-1)}{c}(10c^{7}-26c^{6}-c^{5}-25c^{4}-3c^{3}-27c^{2}-18c-6)\leq 0.
\end{split}
\end{equation*}
One can show that the coefficient of $\log c$ is strictly positive
on $I$: The function $f_{1}(c)=c(100c^{2}-144c+52)$ is positive on
$I$. The function $f_{2}(c)=7c^{6}+12c^{5}-14c^{4}+2$ has two
stationary points on $I$: a local maximum at the origin and a local
minimum at $c_{0}=\frac{-15+\sqrt{813}}{21}$ and $f_{2}(c_{0})>0$.
The coefficient of $\log c$ is
\[96c^{7}+6f_{2}(c)+f_{1}(c)+36c^{3}(1-c)\]
which is positive on $I$.

Taking into account that the coefficients of $\log c$ is strictly
positive on $I$ one can rearrange the terms in the previous
inequality:
\begin{equation}\label{eq:4.1.q}
q(c)=\log(c)+\frac{2(c-1)(10c^{7}-26c^{6}-c^{5}-25c^{4}-3c^{3}-27c^{2}-18c-6)}
      {c(96c^{7}+84c^{6}+156c^{5}-240c^{4}+176c^{3}-144c^{2}+52c+12)}\geq 0.
\end{equation}
We will show this inequality in two steps.
\begin{enumerate}
\item $0<c<\frac{1}{2}$ :
In this case let decrease the function $q(c)$ and show that the inequality
\[q^{*}(c)=\log(c)+\frac{3(c-1)(2c^{7}-1)}
      {c(24c^{7}+24c^{6}+44c^{5}-60c^{4}+44c^{3}-36c^{2}+13c+13)}\geq 0\]
holds. Since $q^{*}\left(\frac{1}{2}\right)>0$, it is enough
to show that $\frac{dq^{*}(c)}{dc}<0$ if $0<c<\frac{1}{2}$.
This follows from the inequality
\begin{equation*}
\begin{split}
\frac{dq^{*}(c)}{dc}=&
  \frac{-1}{c^{2}(24c^{7}+24c^{6}+44c^{5}-60c^{4}+44c^{3}-36c^{2}+13c+3)^{2}}
  \cdot \\
  &\cdot\Bigl( c^{12}(-576c^{3}-1440c^{2}-3216c+2112)
              +c^{10}(-2944c+1472)\\
            & +c^{8} (5296c^{2}-7700c+2640)
              +c^{6} (4800c^{2}-7292c+2508)\\
            & +c^{3} (1800c^{3}-568c^{2}-624c+241)
              +(550c^{3}-441c^{2}+69c+9)\Bigr),
\end{split}
\end{equation*}
where $\frac{dq^{*}(c)}{dc}$ is the sum of six negative functions.

\item $\frac{1}{2}<c<1$ :
Since $q(1)=0$ enough to show that $q'(c)<0$ if $\frac{1}{2}<c<1$.
Let us define the following function
\[\kappa(c)=-c^{2}(24c^{7}+24c^{6}+44c^{5}-60c^{4}+44c^{3}-36c^{2}+13c+3)^{2}
            \cdot q'(c).\]
The aim is to show that $\kappa(c)$ is positive if $\frac{1}{2}<c<1$.
Since
\[\kappa \left(\frac{1}{2}\right)\ ,\
  \kappa'\left(\frac{1}{2}\right)\ ,\
  \kappa^{(2)}\left(\frac{1}{2}\right)\ ,\
  \dots \ ,\
  \kappa^{(9)}\left(\frac{1}{2}\right)>0\]
enough to show that $\kappa^{(10)}(c)>0$ if $\frac{1}{2}<c<1$.
This comes from the equality
\[\kappa^{(10)}(c)=39916800(104832c^{4}+110292c^{3}+50688c^{2}+3963c+1015).\]

\end{enumerate}
\end{proof}

\begin{theorem} \label{th:4.2.}
Part $\beta_{1,k}$ of the scalar curvature is
monotone, that is for given $1>a>b$ positive numbers and eigenvalue $1>\la_{k}>0$
the function
$\beta_{1,k}(a-x,b+x)$
is strictly increasing in the variable
$x\in\left\lbrack 0,\frac{a-b}{2}\right\rbrack$.
\end{theorem}
\begin{proof}
From the identities (\ref{eq:mid}) and (\ref{eq:ing}) it follows, that
\[2\varphi(a,a,\la_{k})+\varphi(a,\la_{k},a)=
  \frac{1}{\la_{k}} \cdot\tau(a/\la_{k}),\]
where
\[\tau(c)=-\frac{2c-3}{2c\log^{2}c}+\frac{1}{c(1-c)\log c}
          +\frac{c}{2(1-c)^{2}}.\]
From these equations one arrives at the expression
\[\beta_{1,k}(a-x,b+x)=\frac{1}{\la_{k}}\cdot \left(
  \tau\left(\frac{a-x}{\la_{k}}\right)+\tau\left(\frac{b+x}{\la_{k}}\right)
  \right).\]
This means that $\beta_{1,k}(a-x,b+x)$ is increasing if
$\tau(\tilde a-\tilde x)+\tau(\tilde b+\tilde x)$ is increasing on
$\lbrack 0,\frac{\tilde a+\tilde b}{2} \rbrack$, where
$\tilde a=a/\la_{k}$, $\tilde b=b/\la_{k}$ and $\tilde x=x/\la_{k}$.
To prove that
\[-\tau'(\tilde a-\tilde x)+\tau'(\tilde b+\tilde x)>0\]
holds, which means that $\tau'(\tilde x)$ decreasing, we will show
that
\[\tau''(\tilde x)<0.\]

We will do this in two steps: we show that the functions
\begin{equation}\label{eq:taus}
\tau_{1}(c)=\frac{3-c}{4c\log^{2}c}-\frac{1}{2c(c-1)\log c},\quad
\tau_{2}(c)=\frac{3-3c}{2c\log^{2}c}-\frac{1}{c(c-1)\log c}+\frac{c}{(1-c)^{2}}
\end{equation}
are concave, and we note that $\tau_{1}(c)+(1/2)\tau_{2}(c)=\tau(c)$.

In the concavity proof of $\tau_{1}(c)$ we work with the function
\[\tau^{*}(c)=-2c^{3}(c-1)^{3}\log^{4}c\cdot \tau_{1}''(c)\]
that is
\begin{equation}
\begin{split}
\tau^{*}(c)=&(6x^{2}-6x+2)\log^{3}x-(x-1)(3x-2)(x-3)\log^{2}x\\
              &+(x-1)^{2}(x^{2}-10x+11)\log x+3(x-1)^{3}(x-3).
\end{split}\nonumber
\end{equation}
To prove the concavity of $\tau_{1}(c)$ enough to show that
$\tau^{*}(c)$ is negative if $c\in I$ and positive if $c>1$.
\smallskip

Since $\lim_{c\to 1}\tau^{*}(c)=0$ enough to show that $\tau^{*(1)}(c)>0$.

Since $\lim_{c\to 1}\tau^{*(1)}(c)=0$ enough to show that $\tau^{*(2)}(c)$
  is negative if $c\in I$ and positive if $c>1$.

Since $\lim_{c\to 1}\tau^{*(2)}(c)=0$ enough to show that $\tau^{*(3)}(c)>0$,
or equivalently
\begin{equation}
\begin{split}
  \rho(c)=&c^{3}\cdot\tau^{*(3)}(c)=(-18c^{3}+36c^{2}-18c+12)\log^{2}c \\
          &+(24c^{4}-138c^{3}+164c^{2}+34c-12)\log c
           +(98c^{4}-276c^{3}-184c^{2}-4c-2)>0.
\end{split}\nonumber
\end{equation}

Using the previous method the positivity of $\rho(c)$ comes from the
positivity of the function $\rho^{(4)}(c)$ and the limits
\[\lim_{c\to 1}\rho^{(0)}(c)=\lim_{c\to 1}\rho^{(1)}(c)=
  \lim_{c\to 1}\rho^{(2)}(c)=\lim_{c\to 1}\rho^{(3)}(c)=0.\]
The limits can be checked.
The positivity of
\[\rho^{(4)}(c)=\frac{1}{c^{4}}\bigl(
 72(8c^{4}-3c^{3}-2c^{2}+c-2)\log c+8(444c^{4}-153c^{3}-50c^{2}+4c+42) \bigr)\]
comes from the series expansions of the $\log$ function. We will show that
another expressions which are less then $\rho^{(4)}(c)$ are positive in two steps.

\begin{enumerate}
\item $c>1$ :
It is known from the calculus that if $c>1$ then
\[\log c>2\frac{c-1}{c+1}.\]
Substituting $2\frac{c-1}{c+1}$ into $\rho^{(4)}(c)$ instead of $\log c$
one arrives at the expression
\[ \frac{1}{c^{4}(1+c)}\cdot
  \bigl(401c^{5}+185c^{3}(c^{2}-1)+93c^{4}+8c(c-1)+78 \bigr) \]
which is positive if $c>1$.
\item $0<c<1$ :
The functions $f_{1}(c)=8c^{4}\log c+1$, $f_{2}(c)=-(3c^{3}+2c^{2}-c+2)\log c$
and $f_{3}(c)=(444c^{4}-153c^{3}-50c^{2}+4c+12)$ are positive on $\lbrack 0,1 \rbrack$.
From this follows that
\[\rho^{(4)}=\frac{1}{c^{4}}\cdot\bigl(72f_{1}(c)+72f_{2}(c)+8f_{3}(c)+312 \bigr)>0
\qquad\text{if}\ 0<c<1.\]
\smallskip

\end{enumerate}

We proved that the function $\tau_{1}(c)$ (was defined by (\ref{eq:taus}))
is concave. In the
concavity proof of $\tau_{2}(c)$ (was defined by (\ref{eq:taus}))
we work with the function
\[\tau^{*}(c)=c^{3}(c-1)^{4}\log^{4}c\cdot \tau_{2}''(c),\]
that is
\begin{equation}
\begin{split}
\tau^{*}(c)=& 2c^{3}(c+2)\log^{4}c-2(c-1)(3c^{2}-3c+1)\log^{3}c
             +(3c-2)(c-3)(c-1)^{2}\log^{2}c \\
           &-(3c^{2}-12c+11)(c-1)^{3}\log c-9(c-1)^{5}.
\end{split}\nonumber
\end{equation}
To prove the concavity of $\tau_{2}(c)$ enough to show that $\tau^{*}(c)$
is positive. Since one can check the following limits
\[\lim_{c\to 1}\tau^{*}(c)=\lim_{c\to 1}\tau^{*(1)}(c)=\dots
  =\lim_{c\to 1}\tau^{*(5)}(c)=0\]
enough to show that the function
\begin{equation}
\begin{split}
\rho(c)=\frac{1}{c^{2}}\tau^{*(5)}(c)=
 &48(2c-1)\log^{3}c+\frac{6}{c^{3}}(100c^{4}-11c^{3}+12c^{2}+12c+12)\log^{2}c\\
 &-\frac{2}{c^{3}}(90c^{5}-246c^{4}-216c^{3}-62c^{2}-9c+78)\log c\\
 &-\frac{1}{c^{3}}(951c^{5}-642c^{4}-403c^{3}-36c^{2}+100c-42)  \\
\end{split}\nonumber
\end{equation}
is positive if $0<c<1$ and negative if $1<c$.
Using the previous methods again one can check that
\[\lim_{c\to 1}\rho(c)=\lim_{c\to 1}\rho'(c)=\lim_{c\to 1}\rho''(c)=0.\]
This means that enough to show that the function
\begin{equation}
\begin{split}
\frac{c^{5}}{2}\cdot\rho^{(2)}(c)=
  &(144c^{4}+72c^{3}+72c^{2}+216c+432)\log^{2}c
  +(-180c^{5}+888c^{4}-78c^{3}-92c^{2} \\
  &-306c-1440)\log c-(1221c^{5}-1056c^{4}+282c^{3}+150c^{2}+273c-870)
\end{split}\nonumber
\end{equation}
is positive if $0<c<1$ and negative if $1<c$.
Since
\[\lim_{c\to 1}\frac{c^{5}}{2}\cdot\rho^{(2)}(c)=0\]
we will show that the function
\begin{equation}\label{eq:eta}
\eta(c)=\frac{-1}{576c^{3}+216c^{2}+144c+216}\cdot
  \frac{d}{dc}\left(\frac{c^{5}}{2}\cdot\rho^{(2)}(c) \right)
\end{equation}
that is
\begin{equation}
\begin{split}
  \eta(c)=&-\log^{2}c+\frac{2(450c^{5}-1920c^{4}+45c^{3}+20c^{2}-63c-432)}
                          {c(576c^{3}+216c^{2}+144c+216)}\cdot\log c\\
          &+\frac{6285c^{5}-5112c^{4}+924c^{3}+392c^{2}+579c+1440}
                 {c(576c^{3}+216c^{2}+144c+216)}
\end{split}\nonumber
\end{equation}
strictly positive.

We will show this inequality in two steps using an appropriate
  approximation of the $\log$ function.

\begin{enumerate}
\item $c>1$ :
It is known that if $c>1$ then $\sqrt{c}>\log c$. If one substitutes
$-c$ into the previous formula instead of $-\log^{2} c$ then one decreases the
$\eta(c)$ function. We decrease again the function and
multiplying by $1152c(11c^{3}+2c+3)$ we show that
\[-(91905c^{4}-9081c^{3}-9064c^{2}+2016c+13824)\log c+
  \frac{(33c^{3}+6c+9)(1855c^{3}-1776c^{2}-72)}{c}\]
is positive. Since the coefficient of the $\log c$ is strictly negative
if $c>1$ enough to show that the function
\[h(c)=-\log c+\frac{3(11c^{3}+2c+3)(1855c^{3}-1776c^{2}-72)}
                   {c(91905c^{4}-9081c^{3}-9064c^{2}+2016c+13824)}\]
is positive if $c>1$. One can check that $h(1)=\frac{3}{800}>0$ and
\[h'(c)=\frac{864(87213719c^{4}+366253c^{3}+339642c^{2}+218160c+10368)+c^{4}q(c)}
             {c^{2}(91905c^{4}-9081c^{3}9064c^{2}+2016c+13824)^{2}}\]
where $q(c)$ is a polynom of $c$. To see that $h'(c)$ is positive
if $1<c$ enough to use the inequalities
\[q(1)\ ,\ q'(1)\ ,\ q^{(2)}(1)\ ,\ q^{(3)}(1)>0\]
and the inequality
\[q^{(4)}(c)=2376(852418875c^{2}-482743225c-4909728)>0\qquad\text{if}\ 1<c.\]

\item $0<c<1$ :
Since $\eta(1)>0$ ($\eta$ was defined by (\ref{eq:eta})) enough to show that
\[36c^{2}(8c^{3}+3c^{2}+2c+3)\cdot\frac{d\eta(c)}{dc}<0.\]
The previous expression is
\begin{equation}
\begin{split}
&(3600c^{8}-1908c^{7}-6876c^{6}-5552c^{5}-20058c^{4}
 +12888c^{3}+3210c^{2}+1080c+1296)\cdot\log c\\
&+(28740c^{8}+4845c^{7}+2991c^{6}+22155c^{5}-35730c^{4}
  -25475c^{3}-7833c^{2}-3933c-3456).
\end{split}\nonumber
\end{equation}
After increasing this expression and dividing by $c^{3}$ we will
show that the function
\begin{equation}
\begin{split}
-2c(954c^{3}+3438c^{2}+2776c+10029)\cdot\log c+(&28740c^{5}+4845c^{4}+2991c^{3}\\
&+22155c^{2}-35730c-25475)
\end{split}\nonumber
\end{equation}
is still negative if $0<c<1$.
Since the coefficient of $\log c$ is negative if $0<c<1$ we will show that
the function
\[\eta^{*}(c)=\log c-\frac{28740c^{5}+4845c^{4}+2991c^{3}+22155c^{2}-35730c-25475}
                         {2c(954c^{3}+3438c^{2}+2776c+10029)}\]
is positive if $0<c<1$. Since $\eta^{*}(1)>0$ enough to show that
\[-2c^{2}(954c^{3}+3438c^{2}+2776c+10029)^{2}\cdot\frac{d\eta^{*}(c)}{dc}\]
is positive. Computing the previous expression we get
\[(472766109c^{2}-59724482c+25548875)+a_{3}c^{3}+a_{4}c^{4}+\dots+a_{8}c^{8}\]
where $a_{3},a_{4},\dots,a_{8}>0$ and the first part is positive if $0<c<1$.

\end{enumerate}

\end{proof}

The monotonicity of the function $\beta_{2,k}(a,b)$ means
that for given positive numbers $1>a>b$ and eigenvalue $1>\la_{k}>0$,
the function
$\beta_{2,k}(a-y,b+y)$
is strictly increasing in the variable
$x\in\left\lbrack 0,\frac{a-b}{2}\right\rbrack$.

Using the $\tilde a=\frac{a}{\la_{k}}$, $\tilde b=\frac{b}{\la_{k}}$ and
$c=\frac{(a+b)}{2\la_{k}}$ notations the monotonicity means that the function
$\beta_{2,k}(c+x,c-y)$ decreasing in the variable $x\in\lbrack 0,c\rbrack$.

Let us define the following functions:
\begin{equation}\label{eq:phi}
\phi_{1}(u)=\frac{1}{u\log u}+\frac{1}{1-u}\quad
  \phi_{2}(u)=\frac{1}{\log u}+\frac{1}{1-u}.
\end{equation}
Using the equalities (\ref{eq:mid})
\[\la_{k}\cdot\beta_{2,k}(a,b)=3w_{\beta}(x,c)+2q_{\beta}(x,c)+2r_{\beta}(x,c),\]
where
\begin{gather}
w_{\beta}(x,c)=\frac{m(c-x,c+x,1)^{2}}{m(c-x,c+x)m(c-x,1)m(c+x,1)},\quad
  r_{\beta}(x,c)=-\frac{m(c-x,1,1)m(c+x,1,1)}{m(c-x,1)m(c+x,1)},\nonumber\\
q_{\beta}(x,c)=\frac{-m(c-x,c-x,c+x)m(c-x,c-x,1)}{m(c-x,c-x)m(c-x,1)m(c-x,c+x)}
         -\frac{m(c+x,c+x,c-x)m(c+x,c+x,1)}{m(c+x,c+x)m(c+x,1)m(c-x,c+x)}.\nonumber
\end{gather}
These function can be computed explicitly:
\begin{align*}
&w_{\beta}(x,c)=\frac{t_{\beta}(x,c)+t_{\beta}(-x,c)}{2},  & &\text{where}\quad
 t_{\beta}(x,c)=\dfrac{\dfrac{\log(c+x)}{\log(c-x)}\cdot\dfrac{c-x-1}{c+x-1}-1}
              {x(\log(x+c)-\log(x-c))}, &  \\
&q_{\beta}(x,c)=q_{1,\beta}(x,c)+q_{2,\beta}(x,c)  & &\text{where}\quad
 \left\lbrace\begin{array}{l}
 q_{1,\beta}(x,c)=\dfrac{\phi_{1}(c+x)-\phi_{1}(c-x)}{\log(x+c)-\log(x-c)},\\[0.5em]
 q_{2,\beta}(x,c)=\dfrac{\phi_{2}(c+x)}{x}+\frac{\phi_{2}(c-x)}{-x},
 \end{array}\right. \\
&r_{\beta}(x,c)=-\phi_{2}(c-x)\phi_{2}(c+x).&
\end{align*}

The monotonicity of the function $\beta_{2,k}$ comes from the inequality
\[\frac{d}{dx}\bigl( 3w_{\beta}(x,c)+2q_{\beta}(x,c)+2r_{\beta}(x,c)\bigr)<0.\]
This inequality is rather difficult to prove but it
can be separated into four seemingly monotone part.

\begin{theorem} \label{th:4.3.}
If for all positive parameters $c$ the following conditions
hold for all $x\in\rbrack 0,c\lbrack$
\begin{align}
&\text{1: }\ \frac{d}{dx}\bigl(w_{\beta}(x,c)+q_{\beta}(x,c)\bigr)<0
&&\text{2: }\ \frac{d}{dx}\bigl(2w_{\beta}(x,c)+q_{1,\beta}(x,c)+r_{\beta}(x,c)\bigr)<0 \\
&\text{3: }\ \frac{d}{dx}\bigl(q_{2,\beta}(x,c)\bigr)<0
&&\text{4: }\ \frac{d}{dx}\bigl(r_{\beta}(x,c)\bigr)<0 \nonumber
\end{align}
then the $\beta_{2,k}$ part of the scalar curvature is
monotone, that is for given $1>a>b$ positive numbers and $1>\la_{k}>0$ eigenvalue
the function
$\beta_{1,k}(a-x,b+x)$
is strictly increasing in the variable
$x\in\left\lbrack 0,\frac{a-b}{2}\right\rbrack$.
\end{theorem}

\begin{lemma} The condition 4 fulfils in the previous theorem.
\end{lemma}
\begin{proof}
The condition
\[\frac{d}{dx}\bigl(r_{\beta}(x,c)\bigr)<0\]
is equivalent to the condition
\[\frac{\phi_{2}'(c+x)}{\phi_{2}(c+x)}>\frac{\phi_{2}'(c-x)}{\phi_{2}(c-x)},\]
where the function $\phi_{2}$ was defined by (\ref{eq:phi}),
so enough to prove that
\[\left(\frac{\phi_{2}'(u)}{\phi_{2}(u)}\right)'>0\]
or equivalently
\[\rho(u)=\phi_{2}(u)\phi_{2}''(u)-(\phi_{2}'(u))^{2}>0.\]
We will show that $\rho_{*}(u)=(1-u)^{4}\log^{4}u\cdot\rho(u)$ is a positive function.
Since
\[\rho_{*}(1)=\rho_{*}^{(1)}(1)=\rho_{*}^{(2)}(u)=\rho_{*}^{(3)}(1)=0\]
enough to show that the function $\tau(u)=u^{4}\cdot\rho_{*}^{(3)}(u)$
that is
\[\tau(u)= 2(-3u^{3}-6u^{2}+u-6)\log^{2}u-2(1-u)(2u^{3}-9u^{2}-3u+2)\log u
          +2(1-u)^{2}(6u^{2}+5u+11)\]
is negative if $0<u<1$ and positive if $1<u$.
Since
\[\tau(1)=\tau^{(1)}(1)=\rho^{(2)}(1)=\tau^{(3)}(1)=\tau^{(4)}(1)=0\]
enough to show that the function
\[\tau_{*}(u)=u^{4}\cdot\tau^{(4)}(u)\]
that is
\[\tau_{*}(u)=(96u^{4}-72u^{3}+48u^{2}+8u+144)\log u
          -8(1-u)(61u^{3}+28u^{2}+28u+30)\]
is increasing or equivalently
\[\tau_{*}^{(1)}(u)=(384u^{3}-216u^{2}+96u+8)\log u
                   \frac{8}{u}(256u^{4}-108u^{3}+6u^{2}+3u+18)>0.\]
After dividing the previous inequality with the coefficient of
$\log u$ which is strictly positive if $u$ is positive, the
inequality is the following:
\begin{equation}\label{eq:d}
d(u)=\log u+\frac{16}{3}+\frac{18}{u}+\frac{1}{3}\cdot
       \frac{2484u^{2}-1284u+655}{48u^{3}-27u^{2}+12u+1}>0.
\end{equation}
We will show this inequality in three steps.
\begin{enumerate}
\item $0<u<\frac{1}{2}$ :
Since $d\left(\frac{1}{2}\right)>0$ enough to check that
$d_{*}(u)=u^{2}(48u^{3}-27u^{2}+12u+1)^{2}\cdot d'(u)$ is negative
if $0<u<\frac{1}{2}$.
Since $d_{*}(0)<0$ enough to show that $d'_{*}(u)<0$ and this inequality
follows from the equation
\[d'_{*}(u)=-u^{5}(25920-16128u)-u^{2}(-37245u^{2}+5400u^{3}+5000)
           -(5260u^{2}-2904u+431),\]
where $d'_{*}(u)$ is a sum of three negative terms.

\item $\frac{1}{2}<u<1$ : One can check that on this interval the
function
\[c(u)=\frac{1}{3}\cdot
       \frac{2484u^{2}-1284u+655}{48u^{3}-27u^{2}+12u+1}\]
is concave hence
\[c(u)>\frac{c(1)-c(1/2)}{1-1/2}\cdot (u-1/2)+c(1/2)\qquad
  \text{if}\ \frac{1}{2}<u<1.\]
Substituting the right hand side of the previous inequality into the
inequality (\ref{eq:d})we get that
\[\log(u)-\frac{113323}{2550}+\frac{18}{u}+\frac{13283}{425}u>0.\]
Let decrease the left hand side of the previous inequality and show that
\[c_{*}(u)=\log u-45+\frac{18}{u}+30u>0\qquad\text{if}\ \frac{1}{2}<u<1.\]
This function has only one stationary point which is a local
minimum on this interval at $u_{0}=\frac{\sqrt{2161}-1}{60}$ and
$c_{*}(u_{0})>0$.

\item $1<u$ :
Since $d(1)>0$ (was defined by (\ref{eq:d})) enough to show that
$c(u)=u^{2}(48u^{3}-27u^{2}+12u+1)^{2}\cdot d'(u)$ is positive if $1<u$.
Since
\[c(1),c'(1),c^{(2)}(1),c^{(3)}(1),c^{(4)}(1)>0\]
enough to show that $c^{(5)}(u)>0$ if $1<u$. This inequality
follows from the equality
\[c^{(5)}(u)=5806080u^{2}-3110400u+893880.\]
\end{enumerate}
\end{proof}

The conditions 1,2,3 in Theorem \ref{th:4.3.} hopefully can be proved similar way.
The numerical evidences for previous theorem is the following:

There was chosen $10^{4}$ parameter $c$ uniformly from the
interval $\lbrack 0,10^{3} \rbrack$ and for each chosen $c$ the
conditions 1,2,3 were numerically checked at $10^{4}$ point
uniformly from the interval $\lbrack 0,c \rbrack$.
This test can be view as a numerical evidence that the
conditions 1,2,3 are valid.
\medskip

The summand $\gamma_{kl}(a,b)$ in expression (\ref{eq:scalkif}) can be computed
using equalities (\ref{eq:mid}), (\ref{eq:ing})
\[\gamma_{kl}(a,b)=\frac{
      3w(\tilde a,\tilde\la_{k})+3w(\tilde b,\tilde\la_{k})+
       q(\tilde a,\tilde\la_{k})+ q(\tilde b,\tilde\la_{k})+
       r(\tilde a,\tilde\la_{k})+ r(\tilde a,\tilde\la_{k})+
       d(\tilde a,\tilde\la_{k})+ d(\tilde a,\tilde\la_{k})}{\la_{l}},\]
where $\tilde a=\frac{a}{\la_{l}}$, $\tilde b=\frac{b}{\la_{l}}$,
$\tilde\la_{k}=\frac{\la_{k}}{\la_{l}}$ and
\begin{align*}
&w(x,c)=\frac{1}{(c-1)\log c}
         \left(\frac{\log c}{\log x-\log c}+\frac{1-c}{x-c}\right)
         \left(\frac{\log c}{\log x}      +\frac{c-1}{1-x}\right),&
&q(x,c)=\phi_{1}(c)\phi_{2} \left(\frac{x}{c}\right), \\
&r(x,c)=\frac{-1}{x}\phi_{2}\left(\frac{c}{x}\right)
                     \phi_{2}\left(\frac{1}{x}\right),&
&d(x,c)=\phi_{2}(c)\phi_{2}(x).
\end{align*}
The function $\phi_{1}(c)$ and $\phi_{2}(c)$ were defined by equation (\ref{eq:phi}).
The monotonicity of summand $\gamma_{kl}(a,b)$ means that the function
$\gamma_{kl}(a-x,b+x)$
is increasing in the variable
$x\in\left\lbrack 0,\frac{a-b}{2}\right\rbrack$.
This property follows from the inequality
\[\frac{d^{2}}{dx^{2}}\bigl(
  3w(x,c)+q(x,c)+d(x,c)+r(x,c) \bigr)<0.\]
This inequality is rather difficult to prove but it seems that
the left hand side of the previous inequality is the sum of two negative
functions.

\begin{theorem} If for all $c$ positive parameters the following conditions
holds for all positive $x$
\[\text{1: }\quad \frac{d^{2}}{dx^{2}}\bigl(2w(x,c)+d(x,c)\bigr)<0
  \qquad
  \text{2: }\quad \frac{d^{2}}{dx^{2}}\bigl(w(x,c)+q(x,c)+r(x,c)\bigr)<0\]
then the $\gamma_{k,l}$ part of the scalar curvature is monotone.
\end{theorem}

The conditions (1) and (2) in the previous theorem were tested numerically
in the parameter space
$x,c\in\lbrack 0,10^{7} \rbrack$, and the computation
confirms the validity of the conditions.
\medskip

Note that equation (\ref{eq:scalkif}) not the only reasonable
decomposition of the scalar curvature formula (\ref{eq:scal}).
There was tested many decomposition and the formula
(\ref{eq:scalkif}) seemed to be the most appropriate one. One can
prove even more about summands in equation (\ref{eq:scalkif}):
\begin{enumerate}
\item $\alpha(a,b)$ (defined by (\ref{eq:alpha})):

For given parameters $a>b>0$ the functions
\begin{align*}
&\va(a-x,a-x,b+x)+\va(b+x,b+x,a-x),   \\
&\va(a-x,b+x,a-x)+\va(b+x,a-x,b+x)
\end{align*}
are strictly monotonously increasing if $x\in\left\lbrack
0,\frac{a-b}{2} \right\rbrack$. From these follows Theorem
(\ref{th:4.1.}).

\item $\beta_{1,k}(a,b)$ (defined by (\ref{eq:beta1})):

For given parameters $1>a>b>0$ and $1>\la_{k}>0$ the functions
\begin{align*}
 &\va(a-x,a-x,\la_{k})+\va(b+x,b+x,\la_{k}), \\
 &\va(a-x,a-x,\la_{k})+\va(b+x,b+x,\la_{k})
            +\va(a-x,\la_{k},a-x)+\va(b+x,\la_{k},b+x)
\end{align*}
are strictly monotonously increasing but the function
\[\va(a-x,\la_{k},a-x)+\va(b+x,\la_{k},b+x)\]
is not increasing if $x\in\left\lbrack 0,\frac{a-b}{2} \right\rbrack$.

\item $\beta_{2,k}(a,b)$ (defined by (\ref{eq:beta2})):

For given parameters $1>a>b>0$ and $1>\la_{k}>0$ the functions
\begin{align*}
 &\va(a-x,b+x,\la_{k})+\va(b+x,a-x,\la_{k}), \\
 &\va(a-x,b+x,\la_{k})+\va(b+x,a-x,\la_{k})
 +\va(a-x,\la_{k},b+x)+\va(b+x,\la_{k},a-x)
\end{align*}
seem to be (numerically tested) strictly monotonously increasing
but the function
\[\va(a-x,\la_{k},b+x)+\va(b+x,\la_{k},a-x) \]
is not increasing if $x\in\left\lbrack 0,\frac{a-b}{2} \right\rbrack$.

\item $\gamma_{k,l}(a,b)$ (defined by (\ref{eq:gamma})):

For given parameters $1>a>b>0$ and $1>\la_{k},\la_{l}>0$ the function
\[\va(\la_{k},a-x,\la_{l})+\va(\la_{k},b+x,\la_{l}) \]
seems to be (numerically tested) strictly monotonously increasing
but the function
\[\va(a-x,\la_{k},\la_{l})+\va(b+x,\la_{k},\la_{l}) \]
is not increasing if $x\in\left\lbrack 0,\frac{a-b}{2} \right\rbrack$.

If one defines another symmetric function
\begin{equation*}
\begin{split}
\gamma_{kl}^{*}(a,b)=& \va(a,\la_{k},\la_{l})+\va(b,\la_{k},\la_{l})+
                       \va(\la_{k},a,\la_{l})\\
                     &+\va(\la_{k},b,\la_{l})+
                     \va(\la_{l},\la_{k},a)+\va(\la_{l},\la_{k},b)
\end{split}
\end{equation*}
then the function $\gamma_{k,l}^{*}(a-x,b+x)$ will be not
increasing if the ratio of $\la_{k}$ and $\la_{l}$ is large enough
($\sim$ 15000).
\end{enumerate}

\section{Scalar curvature on the real and complex state spaces}

The real density matrices form a submanifold of the complex
density matrices. The curvature tensors of a general submanifold
can be very different from the curvature tensors of the manifold.
For example one can think a circle (strictly positive scalar
curvature) as submanifold of the plane (scalar curvature is $0$).
From this point of view one can reformulate Petz's conjecture for
real density matrices. The question arises naturally: is there any
connection between Petz's conjecture for real and complex
matrices.

\begin{theorem} If the scalar curvature is monotone with respect to the majorisation
on the space of complex density matrices then it is monotone on the space of real ones.
\end{theorem}
\begin{proof}
From equations (\ref{eq:scal}), (\ref{eq:scalr}) we get the equation
\[\Scal_{\Bbb R}(D)=\frac{1}{4}\Scal(D)+\frac{1}{4}\sum_{k,l=1}^{n}
  v(\la_{k},\la_{l}),\]
where the function $v(\la_{k},\la_{l})$ was defined by (\ref{eq:ing}).
The monotonicity follows from the next two statements.

1. For every eigenvalues $\la_{k}>\la_{l}$ the function
$v(\la_{k}-x,\la_{l}+x)+v(\la_{l}+x,\la_{k}-x)$
monotone increasing in the variable
$x\in\left\lbrack 0,\frac{\la_{k}-\la_{l}}{2}\right\rbrack$.

2. For every eigenvalues $\la_{k}>\la_{l}$ and $\la_{j}$ the function
$v(\la_{k}-x,\la_{j})+v(\la_{l}+x,\la_{j})+
         v(\la_{j},\la_{k}-x)+v(\la_{j},\la_{l}+x)$
monotone increasing in the variable
$x\in\left\lbrack 0,\frac{\la_{k}-\la_{l}}{2}\right\rbrack$.

Using equalities (\ref{eq:mid}), (\ref{eq:ing}) the function in the
first statement can be written in the form
\[v(\la_{k}-x,\la_{l}+x)+v(\la_{l}+x,\la_{k}-x)=
\frac{1}{\la_{k}+\la_{l}}\left\lbrack (1+c(x))\kappa(c(x))+
  \left(1+\frac{1}{c(x)}\kappa\left(\frac{1}{c(x)}\right)\right) \right\rbrack,\]
where
\begin{equation}\label{eq:kappa}
c(x)=\frac{\la_{k}-x}{\la_{l}+x},\qquad
  \kappa(c)=\frac{3}{2c\log^{2}c}-\frac{2c+1}{c(c-1)\log c}+\frac{c+2}{2(c-1)^{2}}.
\end{equation}
Since the function $c(x)$ is decreasing enough to show that the
function
\[d(c)=(1+c)\kappa(c)+
  \left(1+\frac{1}{c}\kappa\left(\frac{1}{c}\right)\right)\]
is decreasing if $c>1$.
This will follow from the negativity of the function
\[d_{*}(c)=(c-1)^{3}\log^{3}c\cdot d'(c).\]
Since
\[\lim_{c\to 1}d_{*}(c)=\lim_{c\to 1}d'_{*}(c)=0\]
enough to show that
\[\tau(c)=\frac{1}{(1-c)(6c^{4}+18c^{2}+6)}\cdot d''_{*}(c)<0\qquad
\text{if}\ 1<c.\]
Since $\lim_{c\to 1}\tau(c)=0$ enough to show that the function
\begin{equation}
\begin{split}
\frac{3c(c^{4}+3c^{2}+1)^{2}}{(1-c)^{2}}\cdot \tau'(c)=
  &(6c^{6}+22c^{5}+59c^{4}+36c^{3}+59c^{2}+22c+6)\log c\\
  &+\frac{5}{2}(1-c^{2})(c^{4}+6c^{3}+6c+1)
\end{split}\nonumber
\end{equation}
is positive. The coefficient of the $\log c$ is positive, so enough to
show that
\[\tau_{*}(c)=\log c+\frac{5}{2}(1-c^{2})\cdot
  \frac{c^{4}+6c^{3}+6c+1}{59c^{2}+22c+6}<0
   \quad\text{if}\ c>1.\]
Since $\lim_{c\to 1}\tau_{*}(c)=0$ enough to show that $\tau'_{*}(c)$
is positive and this follows from equation
\[ \tau'_{*}(c)=\frac{a_{0}+a_{1}c+a_{2}c^{2}+\dots+a_{12}c^{12}}
{c(6c^{6}+22c^{5}+59c^{4}+36c^{3}+59c^{2}+22c+6)}, \]
where $a_{0},\dots,a_{12}$ are strictly positive numbers.

This completes the proof of the first statement.
\medskip

Using the equalities (\ref{eq:mid}), (\ref{eq:ing}) the function in the
second statement can be written in the following form
\begin{equation}
\begin{split}
v(\la_{k}-x,\la_{j})+v(\la_{l}+x,\la_{j})+
  v(\la_{j},\la_{k}-x)+v(\la_{j},\la_{l}+x)=\frac{1}{\la_{j}}
  \bigl(&\kappa(\tilde\la_{k}-\tilde x)+\kappa(\tilde\la_{l}+\tilde x)\\
        &+\rho(\tilde\la_{k}-\tilde x)+\rho  (\tilde\la_{l}+\tilde x),
\end{split}\nonumber
\end{equation}
where $\tilde\la_{k}=\frac{\la_{k}}{\la_{j}}$,
$\tilde\la_{l}=\frac{\la_{l}}{\la_{j}}$,  $\tilde x=\frac{x}{\la_{j}}$,
the $\kappa(c)$ function was defined by (\ref{eq:kappa}) and
\[\rho(c)=\frac{3}{2\log^{2}c}-\frac{c+2}{(c-1)\log c}+\frac{1+2c}{2(1-c)^{2}}.\]
Using the same arguments as in the proof of Theorem \ref{th:4.2.} enough to
show that the function $\kappa(c)+\rho(c)$ is concave for all positive $c$.
Let us define the function
\[d(c)=\frac{(c-1)^{4}\log^{4}c}{c+5}\cdot (\kappa''(c)+\rho''(c)).\]
To prove the second statement enough to show that $d(c)<0$ for every
positive $c$. Since
\[\lim_{c\to 1}d(c)=\lim_{c\to 1}d'(c)=\lim_{c\to 1}d^{(2)}(c)=\dots=
  \lim_{c\to 1}d^{(6)}(c)=0\]
enough to show that
\begin{equation}\label{eq:taurc}
\begin{split}
\tau(c)=d^{(6)}(c)=&1440\log^{3}c
 -\frac{24}{c^{5}}(30c^{6}-211c^{5}-198c^{4}+207c^{3}+18c^{2}+54c+90)\log^{2}c\\
 &+\frac{8}{c^{5}}(1-c)(351c^{5}+1306c^{4}-29c^{3}+1120c^{2}+957c+765)\log c\\
 &+\frac{4}{c^{5}}(1-c)^{2}(1249c^{4}+1132c^{3}-744c^{2}-872c-705)
  <0\quad\text{if}\ 0<c.
\end{split}
\end{equation}
Note that
\[\lim_{c\to 1}\tau(c)=\lim_{c\to 1}\tau'(c)=0,\quad
  \lim_{c\to 1}(c^{6}\tau'(c))'<0\quad\text{and}\quad
  \tau(4),\tau'(4),(c^{6}\tau'(c))'(4)<0.\]
To prove the inequality (\ref{eq:taurc}) for $0<c<1$ and $4<c$ parameters
enough to check that the function
\begin{equation}\label{eq:taucsillag}
\tau_{*}(c)=(c^{6}\cdot\tau'(c))''
\end{equation}
is positive if $0<c<1$ and negative if $4<c$.
We will prove the inequality (\ref{eq:taurc}) in four steps.
\begin{enumerate}
\item $0<c<\frac{1}{2}$ : After decreasing the function
$\tau_{*}(c)$ and dividing by $c^{2}$ one arrives at the
inequality
\[\psi(c)=-78624\log^{2}c+96\frac{4483c^{2}+108}{c^{3}}\log c-53008c+194856
  -\frac{76944}{c}-\frac{16208}{c^{2}}+\frac{56520}{c^{4}}. \]
Since $\psi\left(\frac{1}{2}\right)=0$ enough to show that
\begin{equation}
\begin{split}
\psi_{*}(c)&=-\frac{c^{4}\psi'(c)}{157248c^{3}+430368c^{2}+31104}\\
           &=\log c+\frac{16}{c}(3313c^{5}-31707c^{3}-2026c^{2}-648c+14130)>0.
\end{split}\nonumber
\end{equation}
Since $\psi_{*}\left(\frac{1}{2}\right)>0$ enough to show that
\[p(c)=3c^{2}(1638c^{3}+4483c^{2}+324)^{2}\psi'_{*}(c)<0.\]
The function $p(c)$ is a polynom and can be checked
that $p(c)<0$ if $0<c<\frac{1}{2}$.

\item $\frac{1}{2}<c<1$ :
After decreasing the function $\tau_{*}(c)$
(which was defined by (\ref{eq:taucsillag})) one arrives at the inequality
\[-819c^{2}\log^{2}c+\frac{4483c^{2}+108}{c}\log c-553c^{3}+2029c^{2}-800c-169
+\frac{588}{c^{2}}>0 \quad\text{if}\ \frac{1}{2}<c<1\]
From the Taylor-expansion of the functions one can check that
\[\frac{4483c^{2}+108}{c}\log>a_{0}+a_{1}(c-1/2)+a_{2}(c-1/2)^{2}, \quad
-819c^{2}\log^{2}c>b_{0}+b_{1}(c-1/2),\]
where
\begin{align*}
&a_{0}=-\frac{4915\log 2}{2},&   &a_{1}=4915-4051\log 2,&
  &a_{2}=3187-864\log 2,\\
&b_{0}=-\frac{819\log^{2}2}{4},& &b_{1}=819\log 2(1-\log 2).&&
\end{align*}
After substituting the $\log c$ functions with the polynom of $(c-1/2)$
we get the following inequality
\[\frac{-1}{4c^{2}}
(2212c^{5}+\alpha_{4}c^{4}+\alpha_{3}c^{3}+\alpha_{2}c^{2}-2352)>0\]
where $\alpha_{4}=3456\log 2-20864$, $\alpha_{3}=3276\log^{2}2+9472\log 2-3712$
and $\alpha_{2}=-819\log^{2}2+4230\log 2+7319$.
One can check that the polynom in the parenthesis is strictly negative if
$\frac{1}{2}<c<1$.

\item $1<c<4$ :
Let increase the function $\tau(c)$ (defined by (\ref{eq:taus})):
substitute $-2500(c-1)$ instead of
\[-\frac{24}{c^{5}}(30c^{6}-211c^{5}-198c^{4}+207c^{3}+18c^{2}+54c+90)\log^{2}c,\]
and $-2000(c-1)^{2}$ instead of
\[\frac{8}{c^{5}}(1-c)(351c^{5}+1306c^{4}-29c^{3}+1120c^{2}+957c+765)\log c \]
and $\frac{4}{c^{5}}\cdot (1-c)^{2}(1249c^{4}+1132c^{3})$ instead of
\[\frac{4}{c^{5}}(1-c)^{2}(1249c^{4}+1132c^{3}-744c^{2}-872c-705).\]
That these substitutions increases the function $\tau(c)$ can be
check using the Taylor-expansion. The new inequality is
\[\tau_{1}(c)=1440\log^{3}c+\frac{4(c-1)}{c^{2}}\cdot
  (500c^{3}-1124c^{2}+117c+1132)<0\quad\text{if}\ 1<c<4.\]
Let increase the function $\tau_{1}(c)$:
\[\tau_{2}(c)=\tau_{1}(c)+2000\frac{(c-1)(c-2)^{2}}{c^{2}}.\]
Since $\lim_{c\to 1}\tau_{2}(c)=0$ to show that $\tau_{2}(c)$ is negative
enough to check that $c\tau'_{2}(c)<0$.
Let increase the function $c\tau'_{2}(c)<0$
\[\tau_{3}=c\tau'_{2}(c)+2000\left(c-\frac{3}{2}\right)^{2}+8(c-4)^{2}-72.\]
Since $\lim_{c\to 1}\tau_{3}(c)=0$ to show that $\tau_{3}(c)$ is negative
enough to check that
\begin{equation}\label{eq:1-4.tau3}
\tau'_{3}(c)=8640\log c+\frac{4(-996c^{4}+608c^{3}+2985c-3472)}{c^{2}}<0
  \qquad\text{if}\ 1<c<4.
\end{equation}
If $2<c<4$ then one can substitute $c-1$ instead of $\log c$ and
one can check that the inequality
\[\frac{-996c^{4}+2768c^{3}-2160c^{2}+2985c-3472}{c^{2}}\quad\text{if}\ 2<c<4 \]
holds.
If $1<c<2$ then one can check the inequality (\ref{eq:1-4.tau3}) holds
using the inequality
\[ \frac{c-1}{c}+\frac{(c-1)^{2}}{2c^{2}}+\frac{(c-1)^{3}}{3c^{3}}
  +\frac{(c-1)^{4}}{4c^{4}}+\frac{(c-1)^{5}}{5c^{5}}+\frac{(c-1)^{6}}{c^{6}}>
  \log c\quad\text{if}\ 1<c<2.\]

\item $4<c$ :
After increasing the function $\tau_{*}(c)$
(which was defined by (\ref{eq:taucsillag})) one arrives at the inequality
\begin{equation}
\begin{split}
(86400c^{3}+59616c+2592)\log^{2}c
  &+\frac{48}{c}(-2985c^{5}+5840c^{4}+3126c^{2}+216)\log c\\
  &+\frac{8}{c^{2}}(2184c^{6}+24357c^{4}+204c+7065)<0\quad\text{if}\ 4<c.
\end{split}\nonumber
\end{equation}
The coefficients of $\log^{2}c$ and $\log c$ are positive and
$c-2>\log^{2}c$, $c-2>\log c$ if $4<c$ therefore one way to
increase the function is to substitute $c-2$ instead of
$\log^{2}c$ and $\log c$. Then the new inequality is
\[c^{4}(5970c^{3}-27948c^{2}+30560c-16855)+(17364c^{3}-216c^{2}+796c-2355)>0
\quad\text{if}\ 4<c.\]
This inequality is holds since the left hand side is the sum of two positive
functions on the $4<c$ interval.
\end{enumerate}
\end{proof}
\bigskip

\section{Conclusions}

We showed that Petz's monotonicity conjecture for the scalar curvature of
  the Kubo-Mori metric follows from more elementary inequalities.
The key idea was to find a good grouping of summands in the expression of the
  scalar curvature.
The proof of some inequality was given using elementary, but brutal computations.
We proved that if Petz's conjecture holds for the manifold of complex
  density matrices then it is also true for the manifold of real ones.

The higher order derivatives of the summands of the scalar curvature play
  central role in these computations.
The scalar curvature is a complicated expression of the Kubo-Mori metric,
  which can be derived from von Neumann entropy.
It seems that these higher order derivatives (of the entropy function)
  responsible for the monotonicity of the scalar curvature.
It would be good to find a proof for Petz's conjecture which is based
  only the easily computable properties of entropy function.
\bigskip
\bigskip

{\bf Acknowledgement.} I would like to thank Dr. D\'enes Petz for
stimulating discussions and valuable remarks. This work was
partially supported by OTKA32374.

\end{document}